*plasma*

# Plasma Medicine: A Brief Introduction

**Mounir Laroussi**

Electrical & Computer Engineering Department, Old Dominion University, Norfolk, VA 23529, USA; mlarouss@odu.edu; Tel.: +757-683-6369

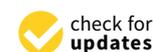

**Abstract:** This mini review is to introduce the readers of *Plasma* to the field of plasma medicine. This is a multidisciplinary field of research at the intersection of physics, engineering, biology and medicine. Plasma medicine is only about two decades old, but the research community active in this emerging field has grown tremendously in the last few years. Today, research is being conducted on a number of applications including wound healing and cancer treatment. Although a lot of knowledge has been created and our understanding of the fundamental mechanisms that play important roles in the interaction between low temperature plasma and biological cells and tissues has greatly expanded, much remains to be done to get a thorough and detailed picture of all the physical and biochemical processes that enter into play.

## 1. Introduction

In the mid-1990s, experiments were conducted that showed that low temperature atmospheric pressure plasmas (LTP) can be used to inactivate bacteria [1]. Based on these results, the Physics and Electronics Directorate of the US Air Force Office of Scientific Research (AFOSR) funded a proof of principle research program in 1997 and supported such research for a number of years. The results from this research program were widely disseminated in the literature, including in peer-reviewed journals and conference proceedings, therefore attracting the attention of the plasma physics community to new and emerging applications of low temperature plasma in biology and medicine [2–8]. The goals of the AFOSR program were to apply low temperature plasmas (LTP) to treat the wounds of injured soldiers and to sterilize/disinfect both biotic and abiotic surfaces. By the early 2000s, research expanded to include eukaryotic cells when small doses of LTP were found to enhance phagocytosis, accelerate the proliferation of fibroblasts, detach mammalian cells without causing necrosis, and under some conditions, lead to apoptosis [9,10].

The above-described groundbreaking research efforts showed that nonthermal plasma can gently interact with biological cells (prokaryotes and eukaryotes) to induce certain desired outcomes. These early achievements raised great interest and paved the way for many laboratories from around the world to investigate the biomedical applications of LTP and by the end of the first decade of the 2000s, a global scientific community was established around such research activities. The field is today known by the term plasma medicine, and in the last few years a number of extensive reviews and tutorials were published (see Refs [11–18] and references therein) as well as a few books [19–21].

Today, the field of plasma medicine encompasses several applications of low temperature plasmas in biology and medicine [22–53]. These include:

- Sterilization, disinfection, and decontamination,
- plasma-aided wound healing

- plasma dentistry
- cancer applications or "plasma oncology,"
- plasma pharmacology,
- plasma treatment of implants for biocompatibility.

In the late 2000s, several LTP sources were approved for cosmetic and medical use. Examples are: in 2008 the US FDA approved the Rhytec Portrait® (plasma jet) for use in dermatology. Also in the US other plasma devices are in use today for various medical applications, such as the Bovie J-Plasma® and the Canady Helios Cold Plasma and Hybrid Plasma™ Scalpel. In Germany, the medical device certification class IIa was given to the kINPen® (plasma jet) in 2013, and the PlasmaDerm® device (CINOGY GmbH) was also approved. Figure 1 is a timeline graph showing the major milestones in the development of the field of low temperature plasma medicine.

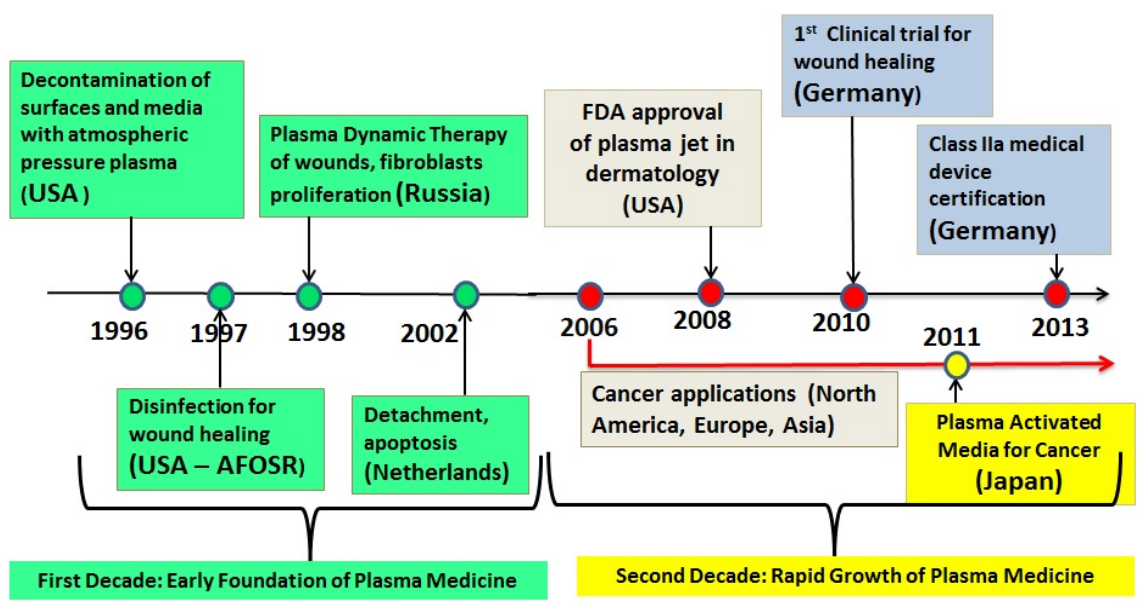

**Figure 1.** Timeline showing some major milestones of the new field of the biomedical applications of low temperature atmospheric pressure plasma. This timeline does not show the case of thermal (hot) plasmas, which were used for many decades in medical applications requiring heat, such as cauterization and blood coagulation.

## 2. LTP Takes on Hygiene and Medical Challenges

As can be seen from Figure 1 the biomedical applications of LTP started with experiments on the inactivation of bacteria on biotic and abiotic surfaces and media. Bacterial contamination proved to pose severe challenges for some industries and in the healthcare arena. The industrial challenges are mainly around the problem of food contamination and sterilization of food packaging. Several well-publicized food poisoning incidents (EHEC, *Listeria*, *Salmonella*) pointed out to consumers that the present methods employed by the food industry may not be adequate to insure food safety. The healthcare challenges are linked to nosocomial infections caused by antibiotic resistant strains of bacteria, such as Methicillin-resistant *Staphylococcus aureus* (MRSA) and *Clostridium difficile* (C-diff). Every year in the US, hospital acquired infections (HAI) kill thousands of patients with compromised immune systems. HAI are caused by inadequate sterilization/decontamination of instruments, surfaces, clothing, bedsheets, and personnel (nurses and doctors). In most cases, contamination by strains of bacteria resistant to the best antibiotic medications available today is the cause of HAI. LTP is therefore considered as a novel technology that can be successfully applied to help solve some of the challenges described above.

The most recent application presently receiving much attention is the use of LTP to destroy cancer cells and tumors in a selective manner [38–58]. Starting around the mid-2000s several investigators reported experiments showing that low temperature plasmas (LTP) can destroy cancerous cells in vitro. This was followed by some in vivo work showing that LTP can reduce the size of cancer tumors in animal models. The in vitro work covered a host of cancerous cell lines, which included glioblastoma, melanoma, papilloma, carcinoma, colorectal cancer, ovarian cancer, prostate cancer cells, squamous cell carcinoma, leukemia, and lung cancer. The in vivo (animal model) work can be found in [38,44,45,53].

In addition to direct plasma applications to cancer cells and tissues, investigators reported that plasma-activated media (PAM) can also be used to destroy cancer cells [38,50,54–58]. Plasma-activated medium is produced by exposing a biological liquid medium to LTP for a length of time (minutes). In this case, the plasma-generated reactive species interact with the contents of the medium and generate solvated long-lived reactive species in the liquid, such as hydrogen peroxide, $H_2O_2$, nitrite, $NO_2^-$, nitrate, $NO_3^-$, peroxynitrite. $ONOO^-$, and organic radicals. These molecules subsequently react with the cells and tissues causing various biological outcomes.

## 3. Mechanisms of Biological Action of LTP: Brief Summary

Investigators reported that the effects of LTP on biological cells (prokaryotes and eukaryotes) are mediated by reactive oxygen and nitrogen species (RONS) [11,12,59–66]. These species include hydroxyl, OH, atomic oxygen, O, singlet delta oxygen, $O_2(^1\Delta)$, superoxide, $O_2^-$, hydrogen peroxide, $H_2O_2$, and nitric oxide, NO. For example, the hydroxyl radical is known to cause the peroxidation of unsaturated fatty acids, which make up the lipids constituting the cell membrane. The biological effects of hydrogen peroxide are mediated by its strong oxidative properties affecting lipids, proteins, and DNA (single and possibly double-strand breaks). Nitric oxide, which acts as an intracellular messenger and regulator in biological functions, is known to affect the regulation of immune deficiencies, cell proliferation, induction of phagocytosis, regulation of collagen synthesis, and angiogenesis.

In cancer cells, the mechanisms of action of LTP are suspected to be related to an increase of intracellular reactive oxygen species (ROS), which can lead to cell cycle arrest at the S-phase, DNA double-strand breaks, and induction of apoptosis. Research by various groups showed that RONS generated by LTP react with cell membranes and can even penetrate the cells and induce subsequent reactions within the cells that can trigger cell-signaling cascades, which can ultimately lead to apoptosis in cancer cells [56–66]. In addition, investigators have shown that plasma-generated RONS can indeed penetrate biological tissues up to depths of more than 1 mm and therefore interact not only with the cells on the surface but with those underneath [67–72].

LTP delivers not only reactive species but it also can exhibit large enough electric fields [73–77]. The magnitudes of these electric fields are several kV/cm and they are suspected to play a role, such as in cellular electroporation, which may allow large molecules to enter the cells.

## 4. Two LTP Sources for Biomedical Applications: Brief Description

The main LTP devices used in plasma medicine research are the dielectric barrier discharge (DBD) and nonequilibrium atmospheric pressure plasma jets (N-APPJ). In fact, the DBD was the device used in the first experiments on the inactivation of bacteria [1]. The DBD uses plate electrodes covered by a dielectric (such as glass). The plasma is generated in the gap separating the electrodes by the application of high sinusoidal voltages in the kHz frequency range. Gases such helium with admixtures of oxygen or air are usually used. For more information on the working of the DBD see references [61,78,79]. Figure 2 shows a schematic of the DBD and a photograph of a diffuse plasma at atmospheric pressure generated by a DBD.

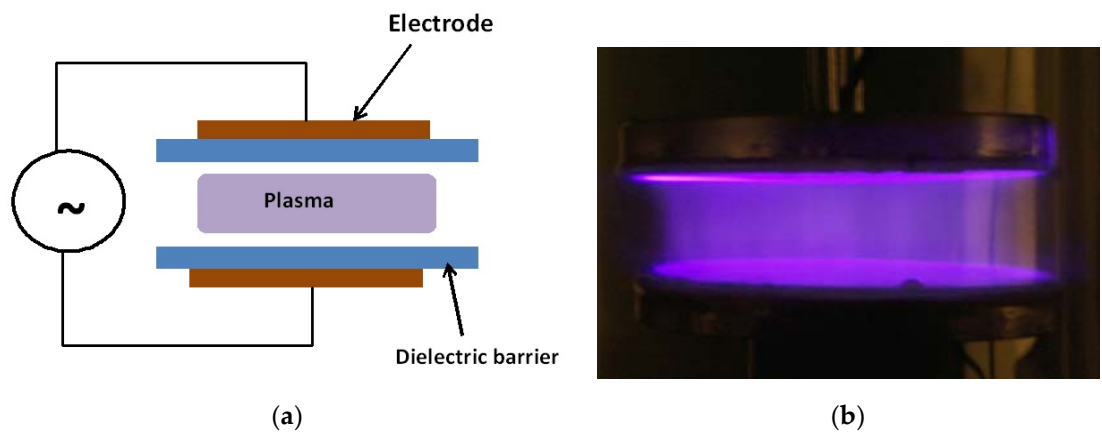

**Figure 2.** Schematic (**a**) and a photograph (**b**) of an atmospheric pressure diffuse plasma generated by a dielectric barrier discharge (DBD). The discharge in the photo on the right is driven by kHz sinusoidal high voltage and the gas is helium with a small admixture of air. Photo taken at the author's laboratory.

Nonequilibrium atmospheric pressure plasma jets (N-APPJs) produce plasma plumes that propagate away from the confinement of electrodes and into the ambient air. The reactive species generated by the plasma can therefore safely and conveniently be transported to a target at a remote location and away from the main plasma generation area. This characteristic made N-APPJs very attractive tools for applications in biology and medicine [60,80–82]. Various power driving methods that include pulsed DC, RF, and microwave power have been used. In addition, various electrode configurations ranging from single electrode, to two-ring electrodes wrapped around the outside wall of a cylindrical dielectric body, to two-ring electrodes attached to centrally perforated dielectric disks have been used. Figure 3 shows photographs of two N-APPJs, the plasma pencil and the kINPen, which have been used extensively in plasma medicine research.

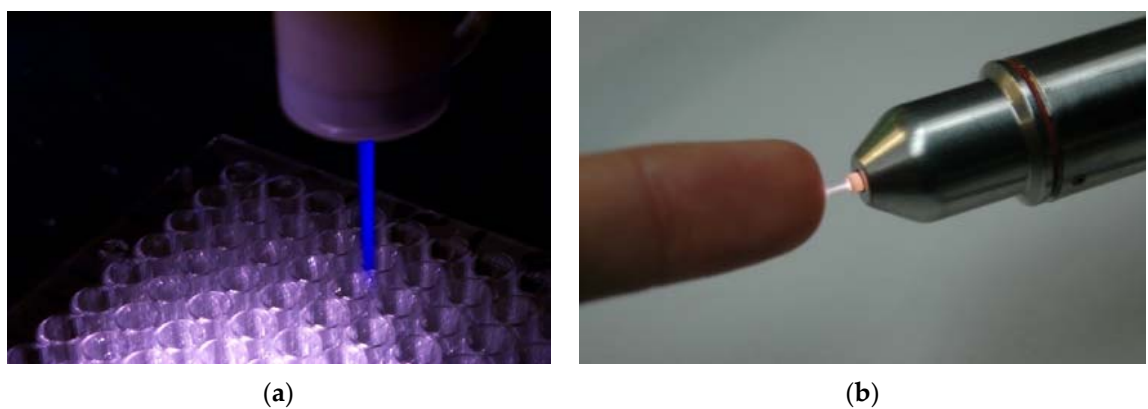

**Figure 3.** Photographs of two plasma jets that have been used in various biomedical applications. (**a**) is the plasma pencil (ODU, Norfolk, VA, USA), and (**b**) is the kINPen (INP, Greifswald, Germany).

The plasma plumes emitted by N-APPJs turned out to be made of small plasma packets traveling at very high velocities (tens of km/s). These plasma packets came to be known as "plasma bullets" and they were independently first reported in the mid-2000s by Teschke et al and by Lu and Laroussi [83,84]. Lu and Laroussi used nanosecond-pulsed DC power while Teschke et al used RF power. The plasma bullets were subsequently researched extensively, both experimentally and by modeling, by various investigators [85–91]. Today there is agreement that the plasma bullets are guided ionization waves. To learn about these guided ionization waves in greater detail, the reader is referred to [92].

## 5. Two Biomedical Applications of LTP

To illustrate the effects of LTP on biological targets, two applications are shown here. The first concerns the bactericidal property of LTP and the second shows the effects of direct plasma exposure as well as plasma activated media on cancerous and healthy epithelial cells. The results presented below are based on the use of the plasma pencil described earlier. The results shown were obtained by the application of the LTP plume generated by the plasma pencil on a bacterial lawn seeded on the surface of a Petri dish (see Figure 4). The bacterium used was *Acinetobacter calcoaceticus*, a gram-negative soil bacterium also found in the tiger mosquito, which is known to be a transmission vector of yellow and dengue fevers. Figure 5 shows zones of inactivation (dark circular areas) around the center of the dish where the plasma plume was applied. The photo to the left is for an initial bacteria concentration of $10^9$/mL, while that on the right is for an initial concentration of $10^7$/mL. It is clear that the killing effects are more extended and pronounced for the lower initial concentration. For more information on the dependence of inactivation on the plasma exposure time and on the type of bacteria, the reader is referred to [93].

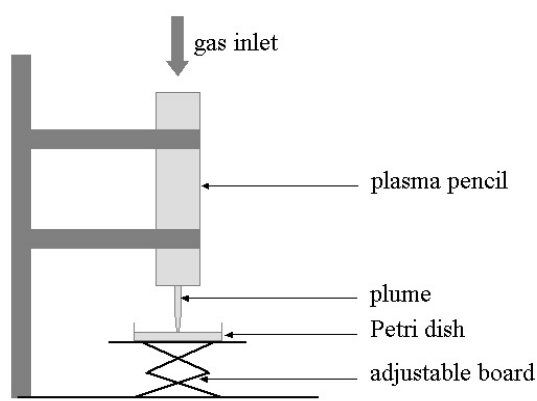

**Figure 4.** Experimental setup for the bacterial inactivation experiments.

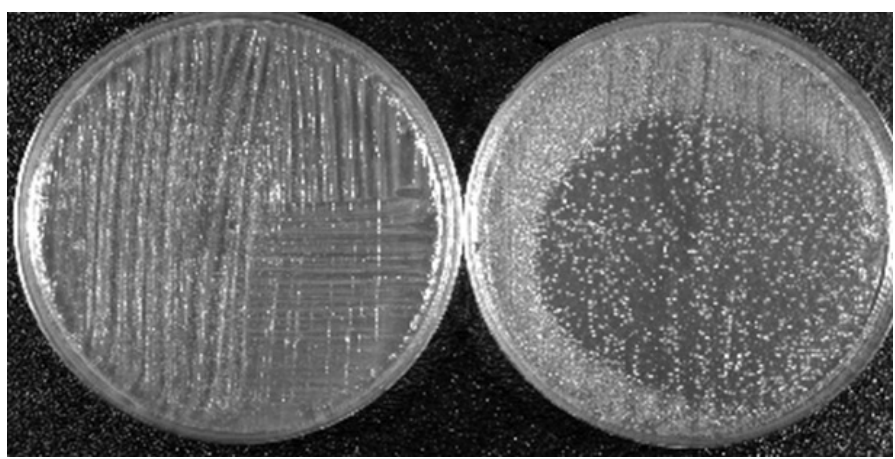

**Figure 5.** Killing property of LTP: Dependence of the killing efficacy on the initial bacteria concentration. Left picture is for $10^9$/mL and right picture is for $10^7$/mL. Bacterium is *A. calcoaceticus*. LTP source is the plasma pencil operated with helium as a carrier gas [93].

Figure 6 shows the effects of direct application of LTP on suspensions of cancerous cells. The cancer cell line used was a squamous cell carcinoma of the bladder (SCaBER, ATCC HTB-3$^{TM}$) originally obtained from a human bladder. After LTP exposure and proper incubation process (37 °C under 5% $CO_2$ atmosphere), Trypan-blue exclusion assay was used to count the number of live and dead cells. For details of the experimental protocol please refer to [40]. The counts immediately after LTP treatment (at 0 h) revealed no dead cells, which suggested there were no immediate physical effects. However, the viability of cells reduced to around 50% at 24 h after a 2-min LTP treatment. As seen in Figure 6, higher plasma exposure times result in more cells killed (5-min plasma treatment results in 75% of loss of viability at 24 h post-treatment) [40]. These results indicate that LTP does not apply immediate brute physical force on the cells, but its effects require longer biological times to show. This is an indication that plasma agents, such as reactive species and electric fields, interact with the cells and induce reactions and/or trigger biochemical pathways that ultimately result in the death of the cancer cells hours later.

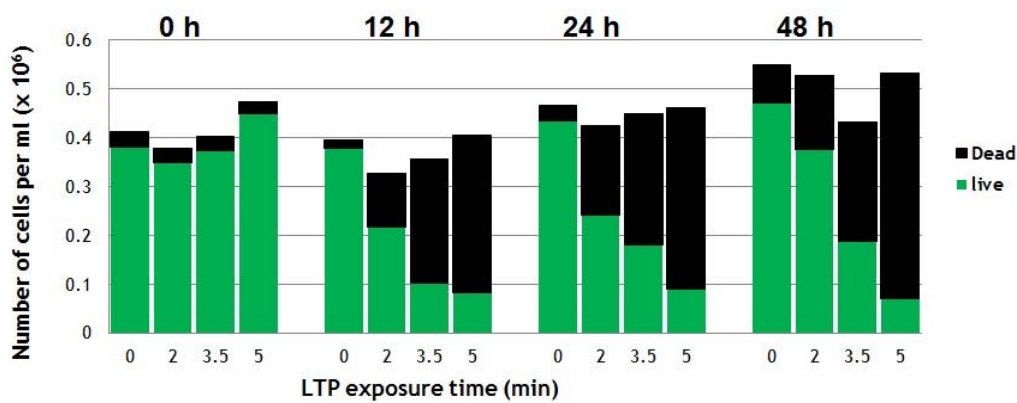

**Figure 6.** Viability of SCaBER cells in media treated directly by the LTP plume of the plasma pencil reveal dead (black bars on top) and live (green bars) cells. The viability was monitored at 0, 12, 24 and 48h post-LTP treatment [40].

Figure 7 shows the selective effect of LTP when it comes to destroying cancer cells versus healthy cells in vitro. The viability results shown in the figure below were obtained using plasma activated media (PAM), which was created by exposing biological liquid media to the plasma pencil for certain lengths of time. The cancerous cell line used was SCaBER and the healthy/normal cells were MDCK (Madin-Darby canine kidney) cells from normal epithelial tissue of a dog kidney. The media used to make PAM were MEM (minimum essential media) for SCaBER and Eagle Minimum Essential Media (EMEM) for MDCK. Figure 7 shows the results [57].

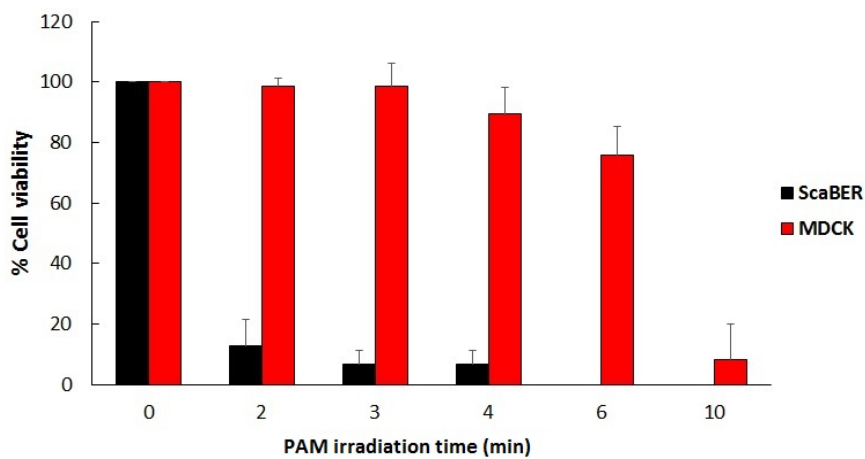

**Figure 7.** Viability in percent of SCaBER (cancerous) and MDCK cells (noncancerous) treated by PAM for various lengths of time. Viability was assessed after 12 hours incubation with PAM using MTS assay and Trypan-blue exclusion assay [57].

Figure 7 shows that PAM created using longer exposures to LTP has increasing killing effects on SCaBER cancer cells, reducing their viability to below 10% for irradiation times longer than 2 min. However, normal MDCK cells were able to withstand exposure to PAM for 3 min. This illustrates the selectivity of PAM in killing cancer cells while sparing healthy cells. But for PAM created with longer exposures to LTP (6 minutes and more) extensive killing of MDCK cells was obtained. This illustrates that the plasma dose is an important factor to take into consideration for optimal outcomes.

## 6. Penetration of RONS in Tissues

One of the key questions in plasma medicine is the following: Do the RONS generated by LTP only interact and affect cells on the surface of a tissue (or tumor) or do they penetrate the tissue and affect cells in deeper layers? Experimental evidence has shown that LTP does indeed affect cells underneath the tissue surface but what remains unclear is how. One possible explanation is what is referred to as the "bystander effect," which implies that there are chemical signals sent by the cells on the surface (in contact with plasma) to cells in the layer below [41]. These signals would trigger reactions similar to those occurring at the cells on the surface, including the onset of apoptosis. However, and to the best of this author's knowledge, there has been no experimental proof this occurs when LTP interacts with tissues. So, the possibility is there, but reliable data that can be replicated needs to emerge first. Therefore, in this section, only experiments that reported qualitatively and/or quantitatively on the penetration of RONS are presented.

In order to qualitatively and quantitatively elucidate RONS penetration into tissues, investigators used various in vitro models. Oh et al. investigated the penetration of RONS using a model made of an agarose film covering a volume of deionized water contained in a quartz cuvette [67]. They found that RONS kept being delivered from the agarose film to deionized water underneath it for up to 25 min after the plasma was removed. To study the delivery of reactive oxygen species (ROS) into cells, Hong et al. used a model comprising phospholipids vesicles encapsulated within a gelatin matrix and equipped with reactive oxygen species (ROS) reporter [68]. They found that ROS were delivered to the cells without rupturing the membranes of the vesicles. To simulate biological tissue, Szili et al. used gelatin gel, a derivative of collagen, and reported on the penetration behavior of $H_2O_2$ through a 1.5 mm thickness gelatin film [69]. The same authors also investigated the effects on DNA in synthetic tissue fluids, tissue, and cells [94].

Tissue models are useful and provide preliminary data regarding the penetration of RONS through biological targets. However, to simulate more realistic conditions, Duan et al. used slices of pig muscle tissue of different thicknesses placed on top of a PBS solution [72]. Figure 8 shows the experimental setup. A plasma jet operated with a helium/oxygen mixture was used. To ignite the plasma sinusoidal high voltages at a frequency of 1 kHz were employed. The plasma treatment times were 0, 5, 10, and 15 min.

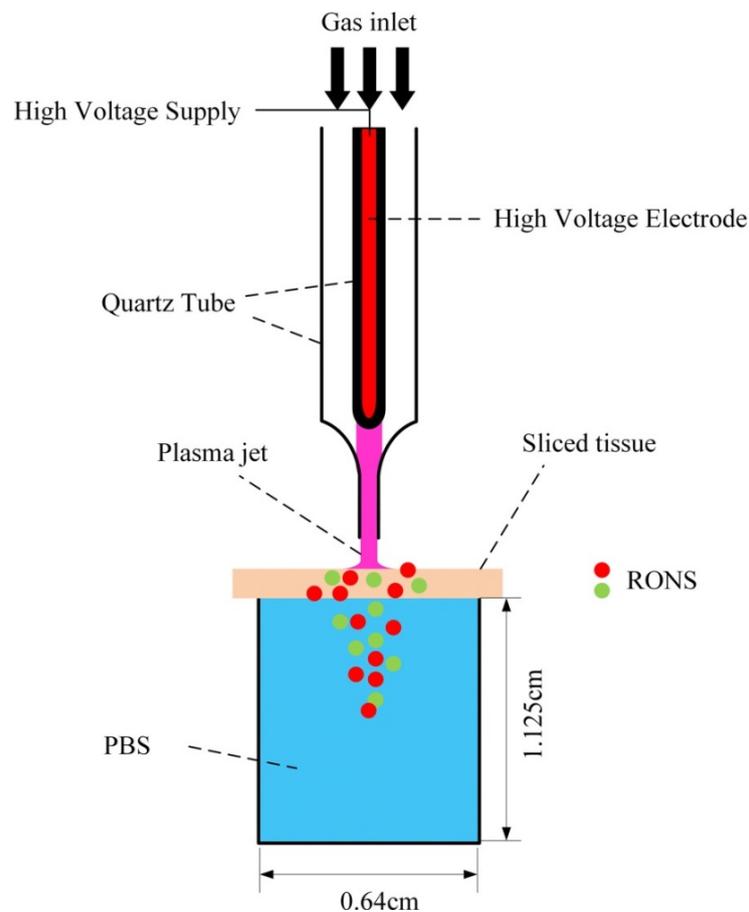

**Figure 8.** Experimental setup using pig muscle tissue [72]. Reproduced from Duan, J.; Lu, X.; and He, G. *Phys. Plasmas* **2017**, *24*, 073506, with the permission of AIP Publishing.

The concentrations of $H_2O_2$, OH, and that of the total of ($NO_2^-$ + $NO_3^-$) were measured for different thicknesses of the tissue slice. A comparison of these concentrations when no tissue was used and when a tissue was placed on top of the solution showed that the concentrations of $O_3$, OH, and $H_2O_2$ were mostly consumed by the tissue and could not pass through 500-µm or greater tissue thickness. However, more than 80% of the ($NO_2^-$ + $NO_3^-$) penetrated a 500-µm-thick tissue slice. Figure 9 shows the measured concentrations of ($NO_2^-$ + $NO_3^-$) as a function of tissue thickness and for three plasma treatment times (5, 10, and 15 min).

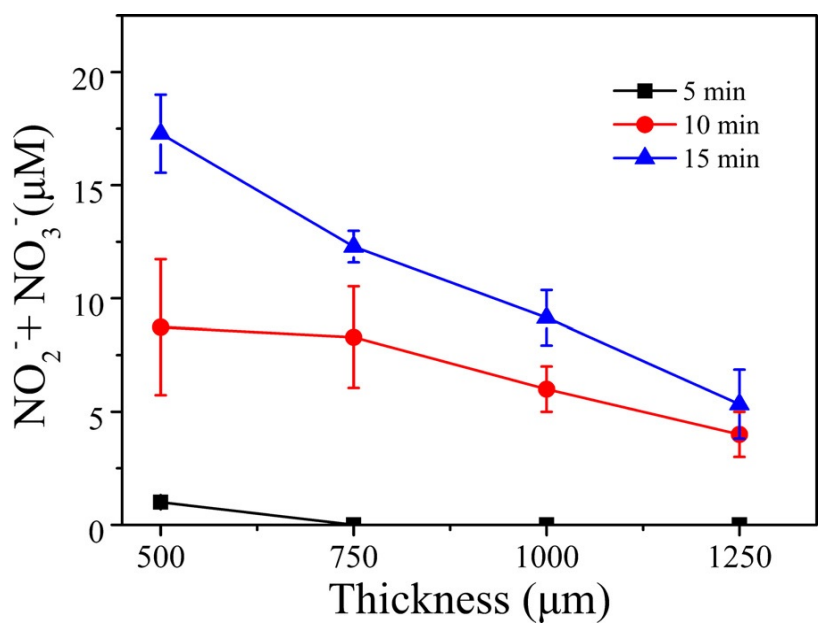

**Figure 9.** Total nitrite and nitrate concentration versus tissue thickness for three plasma exposure times [72]. Reproduced from Duan, J.; Lu, X.; and He, G. *Phys. Plasmas* **2017**, *24*, 073506, with the permission of AIP Publishing.

Figure 9 shows that the concentrations of the nitrogen reactive species, RNS, decrease with the tissue thickness, but increase with the plasma treatment time. The concentration of ($NO_2^-$ + $NO_3^-$) for the 500-μm tissue thickness was comparable to the concentration when no tissue was placed on top of the PBS solution. This means that (RNS) were able to penetrate the tissue slice. This was not the case for ROS, which were absorbed by the tissue, unlike the case when a gelatin model (not real tissue) was used. For that model, ROS were able to penetrate the gelatin film.

The above examples illustrate that RONS do not simply react with the surface of tissues but can indeed penetrate relatively deeply. However, in more realistic conditions using actual tissue, it was shown that not all RONS can cross the same thickness. Some can be absorbed within a few tens of micrometers by the tissue, while others can penetrate up to 1.5 mm below the surface. Of course, the above results may not completely reflect what would happen under in vivo conditions. Such experiments need to be conducted and compared to results obtained for in vitro models and to those obtained under ex vivo conditions [95].

## 7. Conclusions

To conclude this brief introduction of the field of plasma medicine, it is safe to say that the biomedical applications of low temperature plasma have opened up an entirely new multidisciplinary field of research requiring close collaboration between physicists, engineers, biologists, biochemists, and medical experts. This multidisciplinary field started in mid-1990s with seminal experiments on the inactivation of bacteria by low temperature atmospheric pressure plasma generated by a dielectric barrier discharge and slowly expanded to include investigations on eukaryotic cells. Applications in dermatology, wound healing, dentistry, and cancer have led to various scientific advances and to the idea that LTP can be a technology upon which various innovative medical therapies can be developed to overcome present healthcare challenges. However, a lot remains to be done in order to fully understand the mechanisms of action of LTP against biological cells and tissues, both in vitro and in vivo. There is strong indication that LTP acts selectively on cancer cells and tumors and can penetrate deep below the surface, but much more work, including extensive clinical trials, is needed before LTP can be considered a safe technology ready for use in hospitals to treat chronic wounds, cancer lesions and tumors, and other ailments.

**Conflicts of Interest:** The author declares no conflicts of interest.